\documentclass[12pt]{article}
\usepackage{epsfig, bm, amsfonts, amssymb}

\RequirePackage{color}

\title{\bf Universal Amplitude Ratios in the  Ising Model in Three Dimensions}
\author{ 
{\it A. Gordillo-Guerrero$^{\,1,2}$,} {\it R.~Kenna$^{\,3}$,} and {\it J.J. Ruiz-Lorenzo$^{\,4,2}$}\\~\\
$^1$ Departamento de Ingenier\'{\i}a El\'ectrica, Electr\'onica y Autom\'atica,\\
 Universidad de Extremadura,
Avda Universidad s/n, \\
C\'aceres, 10071, Spain. 
{}\\~\\
$^2$ Instituto de Biocomputaci\'on and\\
  F\'{\i}sica de Sistemas Complejos (BIFI),\\
Zaragoza, 50009, Spain.
{}\\~\\
$^3$ Applied Mathematics Research Centre,\\
Coventry University,
Coventry, CV1 5FB, England
{}\\~\\
$^4$ Departamento de F\'{\i}sica,\\
 Universidad de Extremadura,
Avda Elvas s/n, \\
Badajoz, 06071, Spain. 
{}\\~\\
}
\begin{document}
\maketitle
{\Large
  \begin{abstract}
%
We use a high-precision Monte Carlo simulation to determine the 
universal specific-heat amplitude ratio $A_+/A_-$ in the three-dimensional Ising model via the impact angle $\phi$ of 
complex temperature zeros.
We also measure the correlation-length critical exponent $\nu$
from finite-size scaling, and the specific-heat exponent $\alpha$ through hyperscaling.
Extrapolations to the thermodynamic limit yield 
$\phi = 59.2(1.0)^\circ$, $A_+/A_-=0.56(3)$, $\nu = 0.63048(32)$ and 
$\alpha = 0.1086(10)$. 
These results are compatible with some previous estimates 
from a variety of sources and rule out recently conjectured
exact values.
                        \end{abstract} }
%
  \thispagestyle{empty}
%
%
  \newpage
%
                  \pagenumbering{arabic}

\section{Introduction}
\label{I}
\setcounter{equation}{0}

Although it is one of the most investigated models of statistical mechanics,
the still-unsolved Ising model in three dimensions has recently again come
under the spotlight \cite{Ka01,Zh07,Ka11,KlMa08,stDe09,Wu,Perk,MaZh09,MaZh10,LaMa10,FeBl10,Ha10}.  
This is because of (i) controversial claims as to the
exact values of the critical exponents from analytical means and (ii) advances
in algorithmic approaches which have allowed for greater numerical precision.

On the analytic side, two relatively recent papers claim to
have found the exact exponents of the 3D Ising model \cite{Ka01,Zh07}.  They claim  rational values for the critical exponents, including $\alpha = 0$
and $\nu = 2/3$ \cite{Ka11}.  A speculative formula embracing the results for the magnetic
exponent $\delta$ was subsequently given in Ref.\cite{KlMa08} and the conjectured
values for critical parameters in three dimensions were used in Ref.\cite
{stDe09}.  Despite criticism in Refs.\cite{Wu,Perk},
experimental support for these critical exponents  has been claimed
in Ref.\cite{MaZh09}.  Further theoretical arguments in favour of these exponents are
given in Ref.~\cite{MaZh10} (see also Ref.\cite{LaMa10}).  
Refs.\cite{Ka01,Zh07,Ka11} also advocate the existence of a multiplicative logarithmic correction in the critical behaviour 
of the specific heat in the model.

Given the intriguing coincidence of the claimed values of the critical exponents in Ref.\cite{Ka01} and Ref.\cite{Zh07}, 
and their connection with the Rosengren conjecture \cite{Ro86}, precise measurement of $\alpha$ and $\nu$
are of renewed interest.  
Precision is important here, not least
because if any disagreement with RG is found, this will have serious
implications for one of the foundation stones of theoretical physics.  This is
also why the claims of Refs.\cite{Ka01,Zh07} caused so much controversy --
they are not in agreement with precise, though non-exact, results from other
methods.  
Moreover, while the values of the critical exponents claimed in Refs.\cite{Ka01,Zh07} obey
the standard scaling relations, a logarithmic term in the specific heat would
violate the scaling relations for logarithmic corrections unless certain conditions hold for the for the locus of Fisher zeros \cite{KeJo05}. 

Although critical exponents have more commonly been used, amplitude
ratios often offer a more discerning way to identify universality
classes \cite{Izmailian}. These have again recently received increased attention in the case of the of the 3D Ising model because improved numerical techniques now permit more stringent investigations.  The most recent Monte Carlo simulations have yielded the estimates $A_+/A_- = 0.532(7)$ \cite{FeBl10} and $A_+/A_- = 0.536(2)$ \cite{Ha10} and these compare well with previous estimates from a variety of approaches in Table~1.  
The first entry in the table is conspicuous in that (i) it is significantly different to
the other estimates and (ii) it comes from measuring the angle of impact of complex-temperature Fisher zeros onto the critical point, instead of through the more direct approaches employed
by the other references cited.
Therefore we considered it worthwhile revisiting this method, 
making use of the greater computational power available today, 
in an effort to determine whether the inconsistency is indicative of a deeper problem or whether it is simply down to limited numerics.  For these reasons, our analysis focuses on the
scaling and impact angles.
\begin{table}[!b]
\caption{Measurements of the amplitude ratio for the specific heat of the Ising model in three dimensions.
The asterisk indicates that Marinari's Monte Carlo estimate was from the impact angle of Fisher zeros.}  
\begin {center}
\begin{tabular}{l|l|l|l} 
\hline \hline
Authors \& Reference                       & Year   & Method                &    $A_+/A_-$    \\
\hline 
                                           &        &                       &                 \\
Marinari \cite{Ma84}                       & 1984   & Monte Carlo$^\ast$          &   $0.45(7)$     \\
Belanger and Yoshizawa \cite{BeYo87}    & 1987   & Experiment            &   $0.53(1)$     \\
Bagnuls et al \cite{BaBe87}                & 1987   & Field theory          &   $0.541(14)$   \\
Liu \& Fisher \cite{LiFi89}                & 1989   & Series expansions     &   $0.523(9)$    \\
Hasenbusch \& Pinn \cite{HaPi98}           & 1998   & Monte Carlo           &   $0.560(10)$   \\
Nowicki et al  \cite{NoGh01}               & 2001   & Experiment            &   $0.536(5)$    \\
Campostrini et al  \cite{CaPe02}           & 2002   & Series expansions     &   $0.532(3)$    \\
Canfora et al \cite{Canfora}        & 2009   & Phenomonology        &   $0.54$   \\
Feng \& Bl{\"{o}}te \cite{FeBl10}          & 2010   & Monte Carlo           &   $0.532(7)$    \\
Hasenbusch \cite{Ha10}                     & 2010   & Monte Carlo           &   $0.536(2)$    \\
\hline \hline
\end{tabular}
\end{center}
\label{tab1}
\end{table}

Our detailed numerical analysis upholds previous estimates for the 
critical exponents coming from a variety of approaches.
It also brings specific-heat amplitude-ratio estimates
from the impact-angle approach into line with other estimates
and rules out the recent conjectures for exact exponents.

\section{Model and observables}
\label{M}
\setcounter{equation}{0}

The partition function of the pure 3D Ising model 
on a lattice of extent $L$ without an applied magnetic field and at
an inverse temperature $\beta=1/T$ is
\begin{equation}
Z_L(\beta)=\sum_{\{ \sigma_i \}}\mathrm{exp}\left( \beta\sum_{\langle i, j \rangle} \sigma_i \sigma_j \right)\ ,
\label{Z_IM3D}
\end{equation}
where the sum over configurations $\{ \sigma_i \}$ is taken over Ising spins,
$\sigma_i \in \{ \pm 1\}$, and where $\langle i, j \rangle$ denotes nearest neighbours. 
This can also be expressed in terms of the total energy, $E$, as
\begin{equation}
Z_L(\beta)=\sum_{E} p(E,\beta) e^{-\beta E } \,,
\label{ZE_IM3D}
\end{equation}
where $p(E,\beta)$ is the density of states. 
We define the reduced temperature as a dimensionless distance from criticality,
\begin{equation}
 t = 1 - \frac{\beta}{\beta_c} \,,
 \label{reducedt}
\end{equation}
where $\beta_c$ is the critical values of $\beta$. 
The specific heat of the system, is defined as
\begin{equation}
C=\frac{1}{V}(\langle E^2 \rangle_\beta - \langle E \rangle_\beta^2 )\,,
\label{Cesp}
\end{equation}
where $V=L^3$ is the volume of a cubic system of linear extent $L$, and $\langle \ldots\rangle_\beta$ denotes the thermal
average at $\beta$. 
Close to a continuous phase transition, the specific heat scales with reduced temperature $t$ as 
\begin{equation}
 C(t) \simeq A_{\pm} |t|^{-\alpha}\,,
\end{equation}
where the amplitudes $A_+$ and $A_-$ correspond to $t>0$ and $t<0$ respectively.
The susceptibility is also of interest and can be defined as
\begin{equation}
\chi=\frac{1}{V}(\langle M^2 \rangle_\beta - \langle M \rangle_\beta^2 )\,,
\label{susc}
\end{equation}
where $M=\sum_{i} \sigma_i$ is the magnetisation.
One can estimate, for instance, the critical point
using the scaling of the specific heat maxima with the lattice size.
The currently widely accepted value~\cite{PeVi02} for this point is $\beta_c(L=\infty)=0.2216546$.

A complex zero in the partition function indicates a non-analyticity in the free energy.
In the thermodynamic limit ($L\to\infty$), the pinching of such zeros of the real temperature axis precipitates a phase transition.
Nevertheless we can study the transition in a finite system by analytic continuation to complex
temperatures, $\beta=\eta+i\xi$. In this case the partition function includes both
oscillating and damping factors:
\begin{equation}
Z(\beta)=\sum_{E} p(E,\beta)e^{- (\eta+i\xi) E}= \sum_{E} p(E,\beta)
e^{- \eta E} [\cos(\xi E)-i\sin(\xi E)]\ .
\label{ZE_complex3D}
\end{equation}
Rescaling with $Z[\mathrm{Re}(\beta)]$ we define the quantity:
\begin{eqnarray}
R(\eta,\xi) & = & \frac{Z(\beta)}{Z[\mathrm{Re}(\beta)]}=\frac{\sum_{E} p(E,\beta)e^{- \eta E} [\cos(\xi E)-i\sin(\xi E)]}{\sum_{E} p(E,\beta)e^{- \eta E} }
\\
 & = & \langle{\cos(\xi E)}\rangle_\eta-i\langle{\sin(\xi E)}\rangle_\eta
\, .
\label{R}
\end{eqnarray}
Therefore the partition function at complex temperature $\beta=\eta+i\xi$ can be constructed using 
expectation values taken at real temperatures $\beta=\eta$.
A Fisher zero is a complex temperature value $(\eta,\ \xi)$ such that $R=0$. These values can be 
ordered by their distance from $\beta_c(\infty)$ and we denote the first and second zeros by $\beta^{(1)}$ and $\ \beta^{(2)}$,
respectively.

Our objective is to obtain the locations of the Fisher zeros for several system sizes. To achieve this
we performed MC simulations in the vicinity of the critical temperature.
The crucial point is that we can use histogram reweighting techniques~\cite{REWEIGHT} 
to evaluate $\langle{\cos(\xi E)}\rangle_\eta$ and $\langle{\sin(\xi E)}\rangle_\eta$ near the simulation temperature.
With this in hand, we can then minimise $|R|$ using numerical optimisation
methods~\cite{AlDruHans97}.

\section{Finite-Size Scaling of the Fisher Zeros}
\label{FSS}
\setcounter{equation}{0}

In Ref.\cite{IPZ83}  the finite-size scaling (FSS) of the  complex thermal coupling (or Fisher) zeros was obtained.
The zeros  nearest to the real axis $\beta^{(1)}$ scale with lattice size as
\begin{equation}
\beta^{(1)}=L^{-1/\nu} f^{-1}(0)\,,
\label{IPZ1}
\end{equation}
where $f^{-1}(0)$ is a complex number in general. So, both the real and
imaginary parts of the Fisher zeros should scale with the same exponent
$1/\nu$. 
In Appendix A, we discuss an exception of this behavior, and determine under what conditions the real
and imaginary parts scale with different power laws.

In addition there is a relationship between the impact angle of Fisher zeros and the critical amplitudes of the specific heat.
If $\phi$ is the angle between the {\emph{negative}\/} sense of the real temperature-parameter axis and the locus 
of nearby Fisher zeros in the thermodynamic limit, this relationship is
\begin{equation}
\tan\left[ (2-\alpha) \phi \right]=\frac{\cos(\pi
  \alpha)-A_{-}/A_{+}}{\sin(\pi \alpha)}\,.
\label{ratio}
\end{equation}
Notice that if, instead, we define the impact angle $\phi$ as between the {\emph{positive}\/} sense of the real temperature
axis and the locus of zeros then we obtain a similar equation with the roles of $A_{-}$ and $A_{+}$ interchanged:
\begin{equation}
\tan\left[ (2-\alpha) \phi \right]=\frac{\cos(\pi
  \alpha)-A_{+}/A_{-}}{\sin(\pi \alpha)}\,.
\label{ratio2}
\end{equation}
It is clear from either equation that, since the critical exponent $\alpha$
and the amplitude ratio $A_-/A_+$ are universal,  the impact angle is also an universal
quantity. If $\phi=\pi/2$, these equations give that $A_+=A_-$, which is physically reasonable; 
vertical impact signals symmetry on either side of the phase transition and equality of amplitudes.

In this paper, we will follow the first parametrization of the impact angle:
$\phi>0$ will be the angle between the negative direction of the
real temperature axis and the Fisher-zeros locus.
We will then use the relationship (\ref{ratio}) to determine the amplitude ratio
$A_-/A_+$ in a manner independent to \cite{Ma84}, 
and different from Refs.\cite{FeBl10,Ha10,BeYo87,BaBe87,LiFi89,HaPi98,NoGh01,CaPe02}
(see Table~1).
We will show that the different methods yield compatible
results.

\section{Simulation Details}
\label{S}
\setcounter{equation}{0}

\begin{figure}[!ht]
\begin{center}
\includegraphics[width=0.45\columnwidth, angle=270]{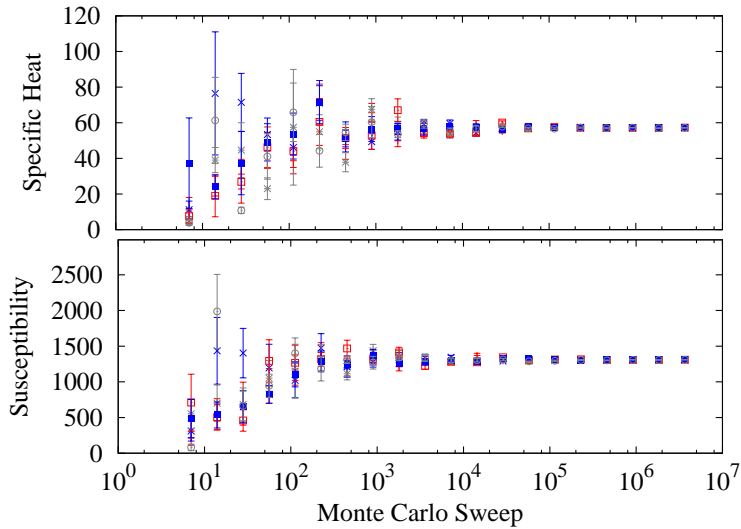}
\caption{(Color online) Log-binning of susceptibility and specific heat for five
random pseudosamples with $L=72$. Error bars are typical deviations inside
each bin. The first block only includes seven measurements, explaining
the deviations for small times.
}
\label{thermatest}
\end{center}
\end{figure}

\begin{table}[b!]
\caption{Simulation details for each system size $L$.
$N_\mathrm{th}$ denotes the number of MCS  (in units of $10^6$) performed in each pseudosample
before taking measurements. We simulated every $L$ at $\beta_c(\infty)=0.2216546$.
In addition, in the third column, we present the second simulation
temperature used for the largest lattices, $\beta'_\mathrm{sim}$.}
\begin{center}
\begin{tabular}{|r|c|c|} \hline \hline
$L$ & $N_\mathrm{th}$   & $\beta'_\mathrm{sim}$   \\\hline
   4       &  1   &  --  \\
   6       &  1   &  --  \\
   8       &  1   &  --  \\
   12      &  1   &  --  \\
   16      &  1   &  --  \\
   20      &  5   &  --  \\
   24      &  5   &  0.22390  \\
   32      &  10  &  0.22300  \\
   48      &  10  &  0.22259  \\
   56      &  30  &  0.22239  \\
   64      &  40  &  0.22213  \\
   72      &  50  &  0.22200  \\
\hline \hline
\end{tabular}
\end{center}
\label{tablesimu}
\end{table}
We performed extensive simulations of the model for linear lattice sizes from
$L=4$ to $L=72$ with periodic boundary conditions.  The spins were updated by
combining ten Wolff single-cluster algorithms with a Metropolis one. We define
in this way our unitary Monte Carlo sweep (MCS). 
We measured after every MCS,
performing $10^7$ measures for each lattice size after thermalisation. We used
a 64-bits pseudo-random generator combining a Parisi-Rapuano
wheel with a congruential one~\cite{PaRa85}.

We simulated twenty independent pseudo-samples for each lattice size starting
from random configurations. We merged their MC histories checking that every
pseudo-sample was thermalised. To check this point we can plot the logarithmic
binning of some quantities for different samples and see the clear
compatibility of the plateaus between samples.
This is illustrated in Fig.~\ref{thermatest} for
our largest system, for both  the specific heat (\ref{Cesp}) and the magnetic susceptibility (\ref{susc}).

We have performed two set of simulations. Firstly we simulated every lattice size
at $\beta_c(L=\infty)=0.2216546$~\footnote{In terms of the associated variable 
$u=\exp(-4 \beta)$, this is given by $u_c=\exp(-4 \beta_c)=0.4120468$.}, 
as this should be the more
direct approach to estimate critical quantities. Nevertheless, we observed unusually
large error bars (always computed via jackknife blocking between samples)
for the locations of the second zeros in the larger lattices. We then performed 
simulations closer to the previously estimated locations of these second zeros for every $L\ge24$. We
call these new temperature set $\{\beta'_\mathrm{sim}\}$. We
obtained fully compatible values for the first zero location from these new temperatures and
we strongly reduced the error bars for the second zeros. It provided us a clear evidence of the sensibility of the
second zero location and discouraged us from studying higher-index zeros. In Table~\ref{tablesimu} we give
the concrete simulation information for each system size. The total estimated simulation time is around a
year of a 3GHz CPU.

\begin{figure}[!t]
\begin{center}
\includegraphics[width=0.45\columnwidth, angle=270]{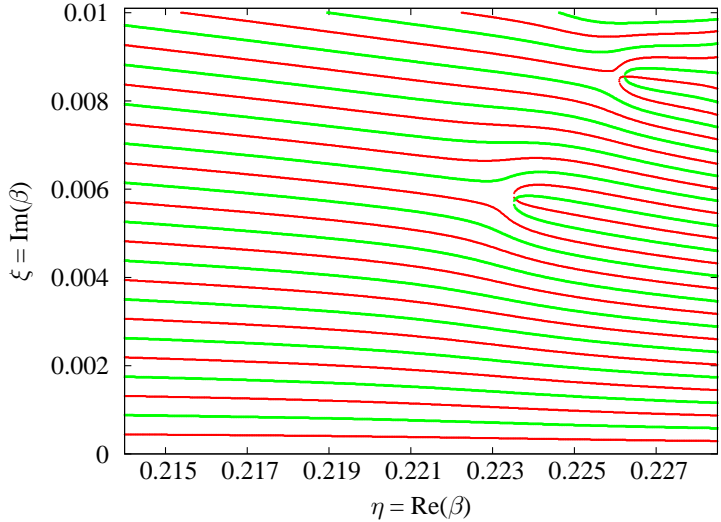}
\caption{(Color online) Graphical analysis for $L=16$ with $2\times10^7$
  measurements and a $1000\times1000$-point extrapolation grid. Dark (red) lines
  represent the points where $\langle{\cos(\xi E)}\rangle_\eta$ changes  sign, light
  (green) lines are the corresponding signals for $\langle{\sin(\xi E)}\rangle_\eta$.
  A crossing of the lines is an imaginary zero of the partition function. The
  extrapolation range allowed the location of the first two zeros.
}
\label{graphical}
\end{center}
\end{figure}

\section{Results}
\label{Ana}
\setcounter{equation}{0}

To determine the values of the universal amplitude ratios we first obtained
the Fisher zeros which lie closest to the real temperature axis for each system size. 
We used the two-step procedure described in Ref.\cite{AlDruHans97}, 
that is: graphical
estimation of the approximate zero location followed by a multidimensional numerical minimization 
starting from the previously estimated point.

For the graphical estimation we had to perform an extrapolation in
$\mathrm{Re}(\beta)=\eta$ (via histogram reweighting, see for example,
Ref.~\cite{FerrSwen}) and $\mathrm{Im}(\beta)=\xi$. Our goal is to obtain the
points where $\langle{\cos(\xi E)}\rangle_\eta$ and $\langle{\sin(\xi E)}\rangle_\eta$ vanish
simultaneously.  We defined the secure $\eta$ extrapolation range using the
square-root of the energy fluctuations as showed in the reweighting
method~\cite{FerrSwen}.  For the optimum $\xi$ range we used the known scaling
of the imaginary parts of the zeros with the system size.  A compromise must
be achieved between resolution (number of extrapolated points) and computation
time. The analysis time for a lattice with $2\times10^8$ measures was around a
week of a 3GHz CPU for a extrapolation grid with $300\times300$ points. In
Fig.~\ref{graphical} we show the output of one of our analyses.

Once we knew the approximate locus of a given zero, we performed an AMOEBA multidimensional minimization~\cite{NUMREC} starting from this point. 
In this way we computed the zeros location with quite high accuracy.  
In Table~\ref{zeros_u} we display the results for the locations of the first two zeros for each lattice size in terms of the conjugate variable $u=\exp{(-4\beta)}$.  
All error bars are computed defining jackknife blocks between the twenty pseudo-samples. 
The larger error bars for the estimations of the second zeros for the smallest lattices ($L\le20$) are due to the longer temperature extrapolation, see table~\ref{tablesimu}.
In order to define the impact angle we use the locus with $\mathrm{Im}\beta >0$ or $\mathrm{Im}u <0$.

\begin{table}[!t]
\caption{First and second partition function zeros in terms of
  $u=\exp{(-4\beta)}$, where $\beta$ is the inverse temperature. Notice that
  if $u$ is a zero its complex conjugate, $u^*$, is also.}
\begin{center}
\begin{tabular}{|r|l|l|l|l|l|} \hline \hline
$L$  & $\mathrm{Re}(u^{(1)})$ & $\mathrm{Im}(u^{(1)})$ & $\mathrm{Re}(u^{(2)})$ & $\mathrm{Im}(u^{(2)})$ \\\hline  
4    &     0.3842660(79)   &   -0.0877447(77)  &    0.344594(055)  &    -0.143306(049)   \\
6    &     0.3975600(88)   &   -0.0454084(75)  &    0.377482(124)  &    -0.072941(165)   \\
8    &     0.4027230(57)   &   -0.0285868(44)  &    0.390136(159)  &    -0.045566(171)   \\
12   &     0.4070150(32)   &   -0.0149382(25)  &    0.400370(224)  &    -0.023291(188)   \\
16   &     0.4088090(27)   &   -0.0094326(32)  &    0.404832(156)  &    -0.014731(191)   \\
20   &     0.4097530(30)   &   -0.0066037(20)  &    0.406870(211)  &    -0.010347(227)   \\
24   &     0.4103170(18)   &   -0.0049403(18)  &    0.408153(030)  &    -0.007684(030)   \\
32   &     0.4109431(15)   &   -0.0031260(20)  &    0.409582(034)  &    -0.004871(034)   \\
48   &     0.4114612(09)   &   -0.0016399(08)  &    0.410759(037)  &    -0.002597(038)   \\
56   &     0.4115855(06)   &   -0.0012821(10)  &    0.411005(022)  &    -0.001995(026)   \\
64   &     0.4116732(07)   &   -0.0010367(07)  &    0.411201(015)  &    -0.001615(022)   \\
72   &     0.4117365(07)   &   -0.0008596(06)  &    0.411343(016)  &    -0.001324(014)   \\
\hline \hline
\end{tabular}
\end{center}
\label{zeros_u}
\end{table}

\subsection{Critical Exponents}

\begin{figure}[!t]
\begin{center}
\includegraphics[width=0.4\columnwidth, angle=270]{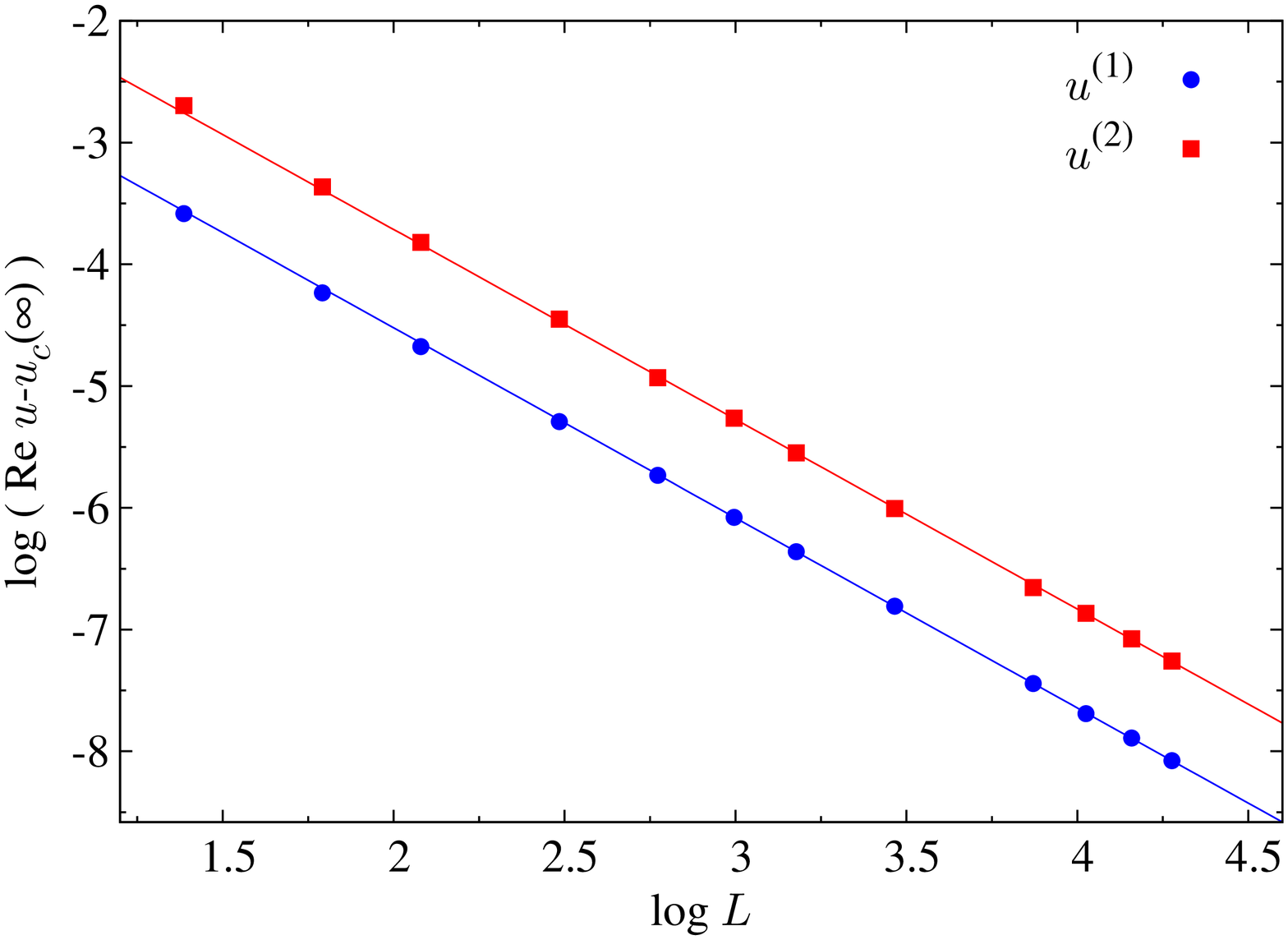}
\includegraphics[width=0.4\columnwidth, angle=270]{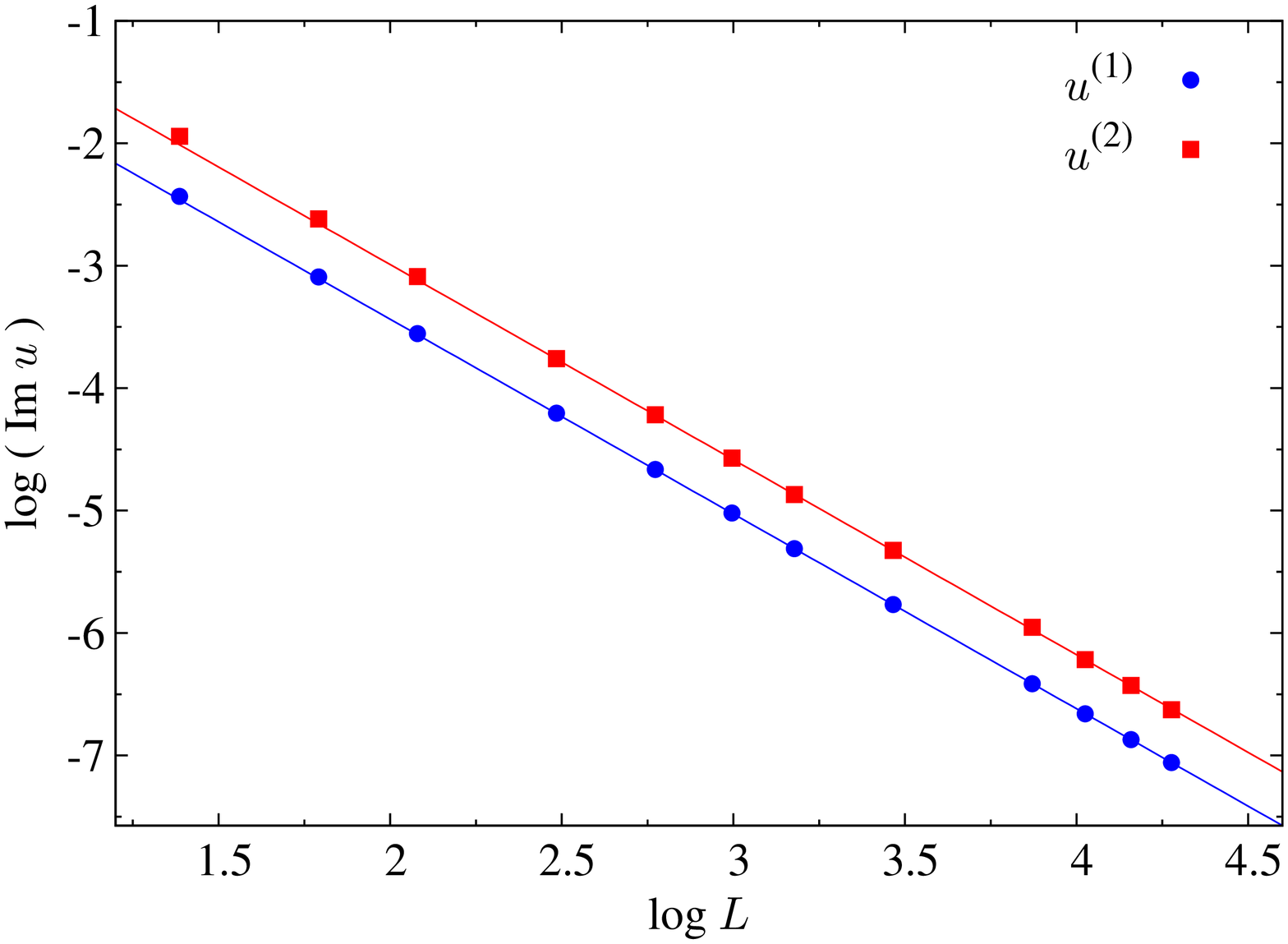}
\caption{(Color online) 
 Top: Scaling of the distance from $u_c(\infty)=0.41204677$
  of the real part of the first (blue circles) and second (red squares)
  Fisher zeros. 
 Bottom: Scaling of the imaginary part of the first (blue
  circles) and second (red squares) Fisher zeros. In each case the error bars 
  are smaller than the symbols.  }
\label{fig_real}
\end{center}
\end{figure}

As was stated in Sec.\ref{FSS}, we can obtain the critical exponent $\nu$ from the scaling of the real and imaginary part of the zeros (see Appendix). 
The scaling of the first two zeros is shown on a double-logarithmic scale in Fig.~\ref{fig_real}. 
As an additional estimator, we can use the distance from the complex zero to the critical point, $|u-u_c|$. 
Therefore we fit each of ${\rm{Im}}( u^{(j)})$,  $u_c - {\rm{Re}}(u^{(j)})$ and $|u^{(j)}-u_c|$ to the form
\begin{equation}
 aL^{-1/\nu}(1+bL^{-\omega})\,,
\label{fit_omega}
\end{equation}
which includes the leading correction-to-scaling term. Firstly we tried to obtain both $\nu$ and $\omega$
from fits of our data sets. We were not able to obtain them simultaneously with the same high accuracy as that in Ref.\cite{Ha10b}. 
Eliminating $\nu$ by taking ratios and fitting 
${\rm{Im}} (u^{(j)})/|{\rm{Re}} (u^{(j)}) - u_c|$ to $ A + B L^{-\omega}$ gives the estimate $\omega = 0.77(9)$ using $L=12$ to $L=72$ or $\omega = 0.63(16)$ using $L=16$ to $L=72$.
These compare reasonably to the estimate $\omega=0.832(6)$ in Ref.\cite{Ha10b}.
Next we decided to focus on the computation of the critical exponent $\nu$. To this end, we fixed the value $\omega=0.832(6)$~\cite{Ha10b}
and obtained the estimates for $\nu$ shown in 
Table~\ref{table_nu}. We discarded data from small sizes until we reached a
good value for the confidence level (CL) of the fit.\footnote{The CL is the probability that $\chi^2$ would be bigger than the observed value, supposing that the statistical model is correct. As a rule, we consider a fit insufficient
 whenever CL$<10$\%.}
To further ensure our control of the corrections to scaling, we checked that our fits are compatible with the fits 
obtained discarding the next smallest $L$ value. We give the error bar corresponding to these latter fits.

Taking our best estimation (the one with the smaller error bar), we obtain the value~\footnote{
Another test of the robustness of our data is to perform
the weighted mean of the data in Table~\ref{table_nu} (where the weights are their inverse variances). In this way we obtain
$\overline{\nu}=0.63069(29)$ and $\overline{\alpha}=0.1079(9)$.}:
\begin{equation}
\nu=0.63048(32)\,.
\label{nu_estimation}
\end{equation}
This is in very good agreement with the world mean value provided in Ref.~\cite{PeVi02} (namely $\nu=0.6301(4)$).
For comparison, the estimate coming from Ref.\cite{Ha10b} is  $\nu= 0.63002(10)$.

\begin{table}[!t]
\caption{Estimated values for the critical exponent $\nu$ from different estimators (first column). The first error bar correspond to the statistical error, while the second one is due to the uncertainty in $\omega$. We give in brackets the size range used in each fit.}
\begin{center}
\begin{tabular}{|r|c|c|} \hline \hline
     & $u^{(1)}$ & $u^{(2)}$ \\\hline
$\mathrm{Re}(u)-u_c$ &  [8--72]: 0.6328(12)(2)   & [4--72]:  0.6576(82)(5)    \\ 
$\mathrm{Im}(u)$     &  [6--72]: 0.63048(25)(7)    & [4--72]:  0.6344(44)(3)    \\ 
$|u-u_c|$            &  [8--72]: 0.63052(41)(2)  & [4--72]:  0.6388(39)(3)    \\
\hline \hline
\end{tabular}
\end{center}
\label{table_nu}
\end{table}

The specific-heat critical exponent can then be estimated using the hyperscaling relation
\begin{equation}
 \nu d = 2 - \alpha\,,
 \end{equation}
 where $d=3$ in the present case. We find 
\begin{equation}
\alpha=0.1086(10)\,.
\label{alpha_estimation}
\end{equation}

The world average coming from Ref.~\cite{PeVi02} is $\alpha =0.1097(12)$ and the estimate from from Ref.\cite{Ha10b} is $0.10994(30)$.

\begin{figure}[!ht]
\begin{center}
\includegraphics[width=0.45\columnwidth, angle=270]{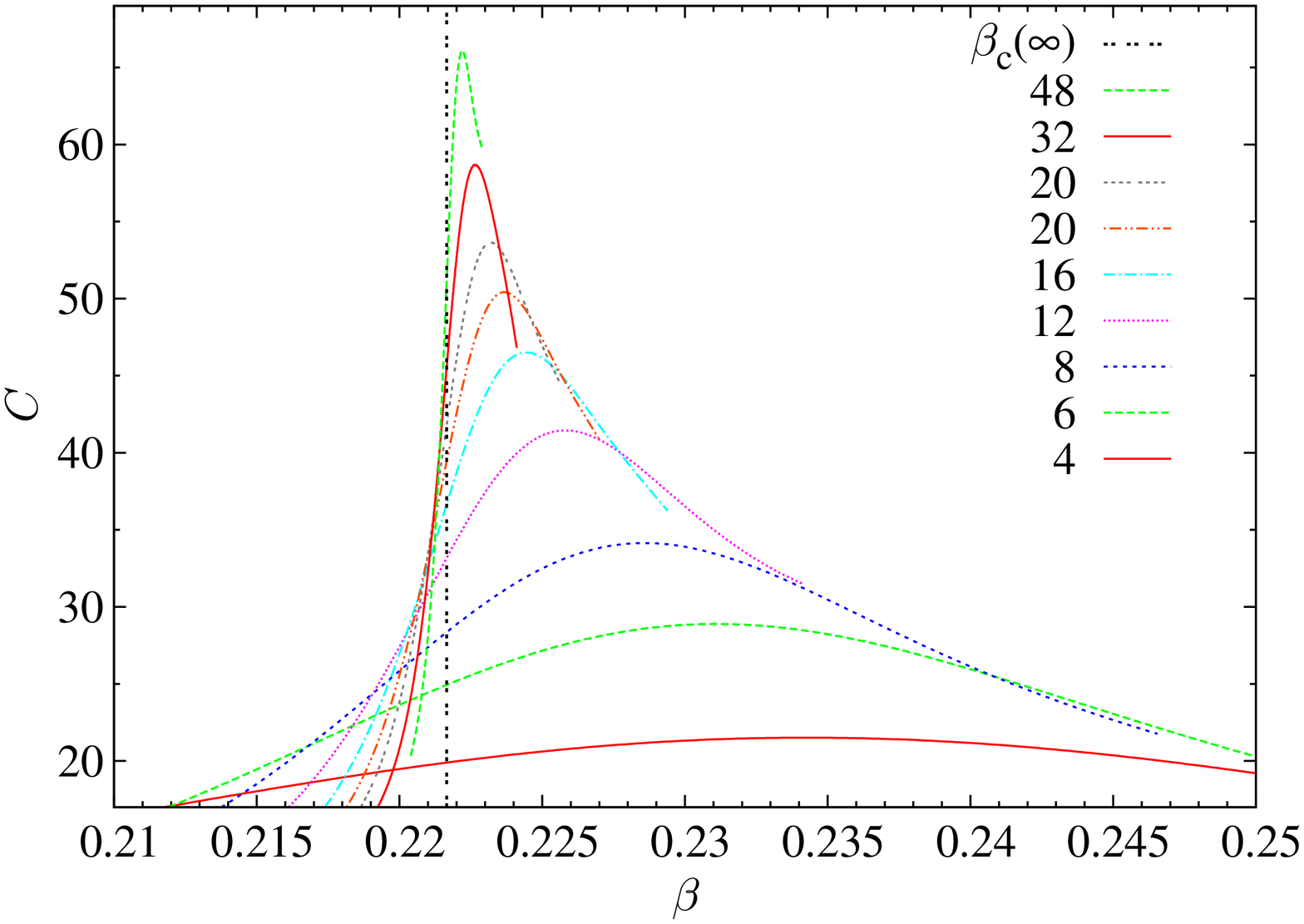}
\caption{(Color online) Specific heat as a function of the inverse temperature
  for different system sizes. We can see the clear approach to the asymptotic
  value $\beta_c(\infty)=0.2216546$ as the system size increases.  }
\label{fig_cesp}
\end{center}
\end{figure}

The conjectures regarding exact values  in Refs.\cite{Ka01,Zh07}
include  a logarithmic divergence in the specific heat.
According to the scaling relations for logarithmic corrections, such a logarithm would necessitate the vanishing of $\alpha$ and
an impact angle other than $45^\circ$ {\cite{KeJo05}.
Since the estimate for $\alpha$ in Eq.(\ref{alpha_estimation}) is over 100 standard deviations away from zero, these conjecture are
 unsupported.

\subsection{Impact Angle and Amplitude Ratios}

\begin{table}[!h]
\caption{
Angles formed between the first and second zeros and the real axis in the $u$ and $\beta$-planes. }

\begin{center}
\begin{tabular}{|r|r|r|r|r|r|r|} \hline \hline
$L$  &$\phi_{1,c}(u)$&$\phi_{1,c}(\beta)$&$\phi_{2,c}(u)$&$\phi_{2,c}(\beta)$ & $\phi_{1,2}(u)$ & $\phi_{1,2}(\beta)$ \\\hline
4    &     72.431(5) &     78.815(5)    &      64.79(2)   &     75.90(2)  &   54.47(5)      &      72.15(5)     \\
6    &     72.31(1)  &     75.55(1)     &      64.64(9)   &     70.1(1)   &   53.9(2)       &      62.6(2)      \\
8    &     71.93(1)  &     73.96(1)     &      64.3(2)    &     67.6(2)   &   53.4(5)       &      58.8(5)      \\
12   &     71.38(1)  &     72.43(1)     &      63.3(5)    &     65.0(5)   &   51.5(9)       &      54.2(9)      \\
16   &     71.05(2)  &     71.72(2)     &      63.9(6)    &     65.0(6)   &   53(2)         &      55(2)        \\
20   &     70.84(2)  &     71.30(2)     &      63(1)      &     64(1)     &   52(2)         &      54(2)        \\
24   &     70.72(2)  &     71.06(2)     &      63.1(2)    &     63.7(2)   &   51.7(4)       &      53.6(4)      \\
32   &     70.54(3)  &     70.76(3)     &      63.2(4)    &     63.5(3)   &   52.1(8)       &      52.6(8)      \\
48   &     70.34(6)  &     70.45(6)     &      63.6(7)    &     63.8(7)   &   54(2)         &      54(2)        \\
56   &     70.38(8)  &     70.47(9)     &      62.4(6)    &     62.6(6)   &   51(2)         &      51(2)        \\
64   &     70.16(6)  &     70.24(5)     &      62.4(5)    &     62.5(5)   &   51(1)         &      51(1)        \\
72   &     70.13(4)  &     70.19(3)     &      62.0(6)    &     62.1(6)   &   50(2)         &      50(2)        \\
\hline \hline
\end{tabular}
\end{center}
\label{tab_cesp}
\end{table}

We also considered it interesting to contrast the location of the Fisher zeros with the temperatures where the specific heat is maximum. 
For every lattice size, we extrapolated Eq.~(\ref{Cesp}) from the simulated temperature to obtain its maximum, see Fig.~\ref{fig_cesp}.
In Fig.~\ref{fig_angle} we plot the locations of these maxima and the locations of the Fisher zeros for every lattice size. 
From this plot one can see that the angle between the positive sense of the real axis,
the specific-heat pseudocritical point and the first Fisher zero change significantly as the lattice size increases.

\begin{figure}[!t]
\begin{center}
\includegraphics[width=0.45\columnwidth, angle=270]{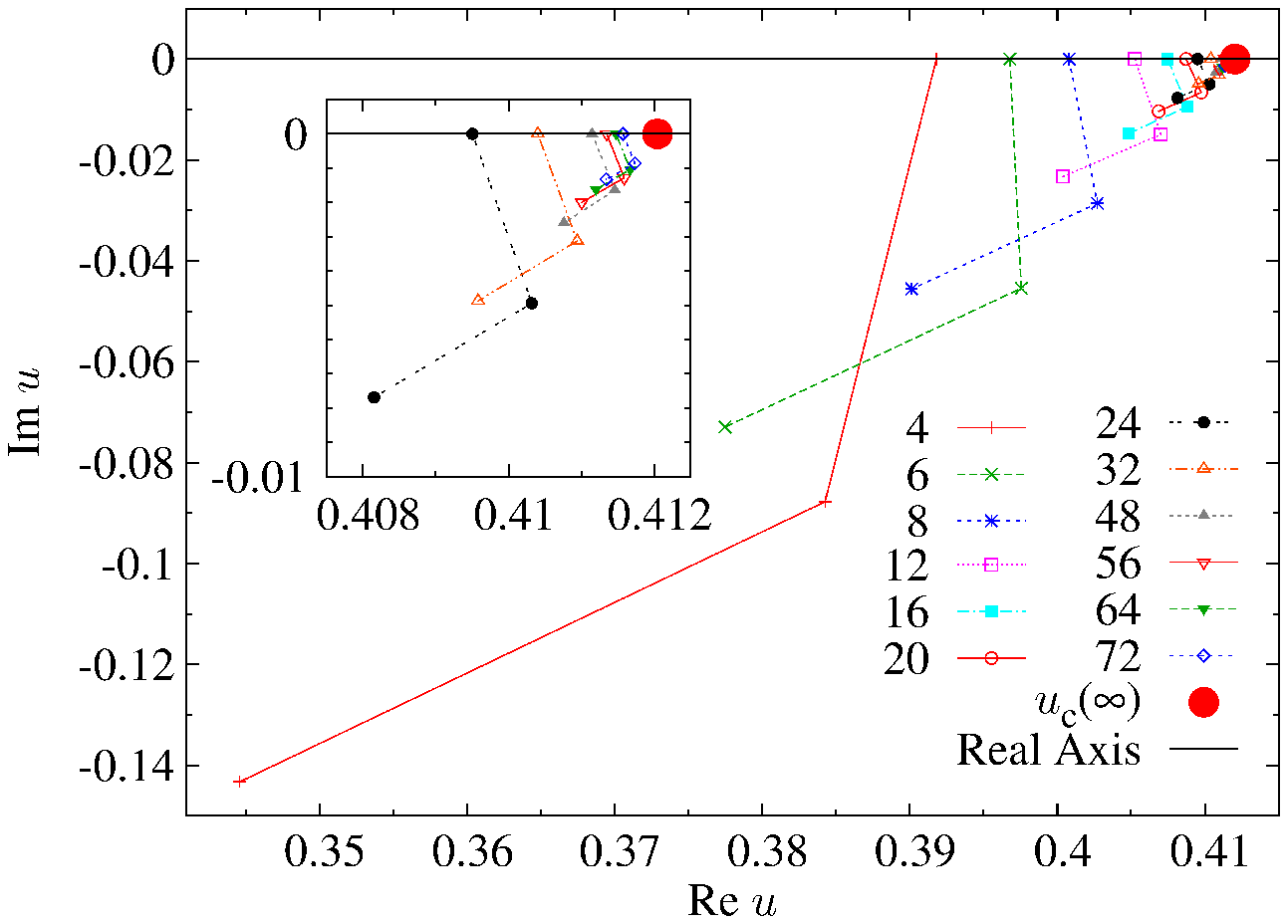}
\caption{(Color online)
First two zeros  as a function of the system size. 
The locations of the pseudocritical temperatures from the maximum of the specific heat are also given on real axis.
We indicate with a large red disc the asymptotic value $u_c(\infty)$. 
The inset is a  magnification for larger lattices.}
\label{fig_angle}
\end{center}
\end{figure}

There are very few works in the literature which use the impact angle of the Fisher zeros to determine amplitude ratios.
Those which do exist are rather old and only had access to very small lattice sizes. 
Marinari's estimate~\cite{Ma84} $A_+/A_-=0.45(7)$ is based on lattices of extent $L=4$ to $L=8$.
An additional motivation for our investigations is to re-examine the impact angle in the light of the much larger 
lattices available to us today.

Unfortunately, there is no FSS theory for the impact angle for Fisher 
zeros\footnote{
One could attempt to build a FSS theory for the impact angle by noting that,
in the infinite-volume limit, one may describe the locus of zeros as
$ u(r) = u_c + r\exp{(i \phi(r))}$.
Taylor expanding  $\phi(r) = \phi + r \phi^\prime  + \dots$,
 one finds that the impact angle is given by
\begin{equation}
 \tan{\phi} = \frac{{\rm{Im}}u(r)}{{\rm{Re}}u(r)-u_c} - r \phi^\prime \sec^2{\phi} + {\cal{O}}(r^2).
\label{tan}
\end{equation}
To achieve the known FSS of the zeros $u^{(j)}(L)$ \cite{IPZ83}, write
$
 r_j(L) = R \left({{j}/{L^d}}\right)^\frac{1}{\nu d}.
$
Replacing $r$ by $ r_j(L)$, 
\begin{equation}
 \tan{\phi} = \tan{\phi_j(L)} - \rho_j L^{-\frac{1}{\nu}} + \dots ,
\label{candidate}
\end{equation}
where
$
 \phi_j(L) = {{\rm{Im}}u^{(j)}(L)}/{({\rm{Re}}u^{(j)}(L)-u_c)}
$
and   $u^{(j)}(L) \equiv u(r_j(L))$ and 
 $\rho_j = R j^{1/\nu d} \phi^\prime \sec^2{\phi}$.
Unfortunately we have found this approach fails due to strong corrections.
}}. 
However, a number of ways to estimate the impact angle suggest themselves. 
These include the angle at the critical point between  the negative sense of the real axis and the 
first  or second Fisher zero as well as the angle formed by the impact of the line joining these
zeros and the real axis. 
In Table~\ref{tab_cesp}, each of these are listed in both the $u$- and $\beta$-planes.
The notation is as follows:
\begin{itemize}
 \item $\phi_{j,c}(u) = $ angle at $u_c$ between the $j^{\rm{th}}$ zero and the real $u$-axis.
 \item $\phi_{1,2}(u) = $ angle between the line joining $u^{(1)}(L)$ and $u^{(2)}(L)$ and the real $u$-axis.
 \item $\phi_{j,c}(\beta) = $ angle at $\beta_c$ between the $j^{\rm{th}}$ zero and the real $\beta$-axis.
 \item $\phi_{1,2}(\beta) = $ angle between the line joining $\beta^{(1)}(L)$ and $\beta^{(2)}(L)$ and the real $\beta$-axis.
\end{itemize} 
Angles measured in the $u$- and $\beta$- planes should converge since the transformation from $\beta$ to $u$ is a conformal one, which preserves angles locally. 
However, for small lattices, where the zeros are relatively far apart, we may expect different impact angles for the $\beta$- and $u$-planes.
But in the large $L$ region and when the zeros are sufficiently close to the real axis, the conformality of the transformation implies that both angles should be nearly equal. 
This fact serves as a check that we are approaching the asymptotic region in the computation of the impact angle
and we can safely discard the results from lattices that are too small to achieve approximate conformality. 
We find that approximate conformality kicks in at around $L \gtrsim 32$.
This means that the earlier measurements of Refs.\cite{Ma84,IPZ83} are unreliable and justifies our re-visitation of the problem.
The results for the various angles and the various lattice sizes are given in Table~\ref{tab_cesp}. The angles were obtained
for the data simulated at $\{\beta'_\mathrm{sim}\}$ to use the correlations between
the two zeros to reduce the errors via the jackknife method.

In the absence of a FSS theory for impact angles, we conservatively take  the wedge formed by the 
smallest and largest angles for which the lattices exhibit approximate conformality to approximate
the approach of the zeros to the real axis. This wedge can be used to estimate upper and lower bounds for the
impact of the locus of Fisher zeros onto the real axis. The impact angle may then be approximated by the 
average of these bounds, with their difference delivering an error estimate. 
In Table~\ref{tab_angle_averaged}, these averages are listed for lattice sizes that exhibit the 
desired approximate conformality.

From the values listed in Table~\ref{tab_angle_averaged}, we 
observe a drift towards smaller values of $\phi$ 
as the lattice size increases. The values for the largest lattice
studied ($L=72$) and its associated estimate for the amplitude ratio are
\begin{equation}
 \phi = 59.9(8)^\circ\,, \quad \quad \quad \frac{A_+}{A_-} = 0.58(2)\,.
\label{final72}
\end{equation}
On may alternatively
attempt to extrapolate to the thermodynamic 
limit using a correction ansatz\footnote{The rationale for this fit is the
  following. We know that
$$
\tan \phi(L)=(\mathrm{Im}(u^{(2)})-\mathrm{Im}(u^{(1)}))/(\mathrm{Re}(u^{(2)})-
\mathrm{Re}(u^{(1)}))\, .
$$ 
The scaling of the numerator and denominator is given by
eq. (\ref{fit_omega}), and finally assuming small angles we obtain
$$
\phi=\phi(L)+O(L^{-\omega})\,.
$$
}
\begin{equation}
  \phi(L)=\phi + b L^{-\omega}\,,
\label{fit_phi}
\end{equation}
where $\phi(L)$ is the impact angle computed with the data from the lattice
size $L$ and $\phi$ is the asymptotic value of this angle.
 
With $\omega = 0.832(6)$ from (\ref{nu_estimation}), we obtain (fitting only $L \ge
48$)  
\begin{equation}
 \phi = 59.2(1.0)^\circ\,.
\end{equation}
The corresponding estimate for the amplitude ratio is
\begin{equation}
 \frac{A+}{A_-} = 0.56(3)\,.
\end{equation}
This compares with Hasenbusch's estimate $A_+/A_{-}= 0.536(2)$
from Ref.\cite{Ha10} (see also  Table~\ref{tab1}).

\begin{table}[!b]
\caption{Impact angle estimation from the average between $\phi_{1,c}(u)$ and $\phi_{1,2}(u)$
only for the sizes respecting conformality.
We also estimate the universal amplitude ratio $A_+/A_-$ from this angle.
}
\begin{center}
\begin{tabular}{|r|c|c|c|c|} \hline \hline
$L$  & Angle Average & $(A_+/A_-)_\mathrm{average}$\\\hline
32     &  61.3(4)     &     0.610(9)   \\
48     &  62(1)       &     0.63(2)    \\
56     &  60.5(8)     &     0.59(2)    \\
64     &  60.5(7)     &     0.59(2)    \\
72     &  59.9(8)     &     0.58(2)    \\
\hline \hline
\end{tabular}
\end{center}
\label{tab_angle_averaged}
\end{table}

\section{Conclusions}
\label{Con}
\setcounter{equation}{0}

Because of recent claims around the values of the critical exponents of the
three-dimensional Ising model, and in the light of renewed interest in the
measurements of amplitude ratios in this model, we decided to investigate this
well-studied model from an alternative perspective, namely that of
complex-temperature zeros. In particular, the manner in which such zeros
impact onto the real axis dictates the amplitude ratios for the specific
heat. Only very old (almost 3 decades old) measurements of this impact angle
exist in the literature, so another motivation was to investigate this using
the powerful algorithms and large lattices currently at our disposal.  Indeed,
we have found that previous measurements of the impact angle were made on
lattices which were too small to manifest conformality in the complex
temperature plane.

In the absence of a finite-size scaling theory for the impact angle, we were
able to estimate upper and lower bounds for its value. These lead to an
estimate $\phi = 59.2(1.0)^\circ$, which in turn leads to an estimate for the
amplitude ratio of $A_+/A_- 0.56(3) $.  In addition to the amplitude ratios,
we investigated the correlation-length critical exponent and determined that
$\nu = 0.63048(32)$.  The corresponding value for the specific-heat critical
exponent is $\alpha = 0.1086(10)$, from hyperscaling.  These values are
compatible with the most accurate recent measurements, and therefore
reinforce confidence in our estimates for the impact angle.  All of our
estimates are consistent with the values currently widely accepted and are
incompatible with the values recently claimed in Refs.\cite{Ka01,Zh07}.

\vspace{1cm}
\noindent
{\bf{Acknowledgements:}} RK acknowleges support from EU Programme
FP7-People-2010-IRSES (Project No 269139), and the ARF Scheme of Coventry
University. AGG and JJR acknowlege support from Research Contracts
No. FIS2007-60977 (MICINN), GR10158 (Junta de Extremadura) and ACCVII-08
(UEX).

\appendix

\section{Appendix}

In FSS theory, the shift exponent $\lambda$ characterizes the scaling of the
pseudocritical point towards the critical point as the thermodynamic limit is
approached. The pseudocritical point is given by the location of the
specific-heat maximum or by the real part of the first Fisher zero and the
approach is as $L^{-\lambda}$.  In most models exhibiting higher-order phase
transitions, the shift exponent $\lambda$ coincides with the inverse of the
correlation-length critical exponent $1/\nu$, but this is not a direct
conclusion of FSS theory and is not always true.

For example, for the Ising model in two dimensions, Ferdinand and Fisher determined that the specific-heat
pseudocritical point scales with $\lambda = 1/\nu = 1$ \cite{ferdinand1969}. 
However, Ising models defined on other two-dimensional lattices with different topologies have
shift exponents which differ from the inverse correlation length
critical exponent (see Ref.\cite{JaKe02} for a discussion and references therein).
This is despite the fact that the critical properties on such lattices are the same as for the torus in the thermodynamic limit.
Here we ask the question: Is there a criterion for  $\lambda \ne 1/\nu$?
We find that there is such a criterion --  the specific-heat amplitude ratio must be $1$.

Although a detailed FSS theory for the impact angle is lacking, for sufficiently large $L$ one may expect that 
$\phi$ is approximated by  the angle subtended by the first zero on the real axis at the critical point,
  \[
   \tan \phi \approx \frac{{\rm{Im}} [u^{(1)}(L)]}{{\rm{Re}}[ u^{(1)}(L)] - u_c}
    \propto \frac{L^{-1/\nu}}{L^{-\lambda}} \sim C L^{\lambda - 1/\nu} + \dots.
    \]
To find the angle $\phi$, let $L \rightarrow \infty$.
 If, as in most cases, $\lambda = 1/\nu$, then $\tan{\phi} = C$, some constant value.
 If $\lambda < 1/\nu$ then $\tan \phi = 0$. But this is impossible as $\lambda$ cannot be less than $1/\nu$ (otherwise $\nu$ \emph{becomes} $1/\lambda$).

The only possibility to have $\lambda \ne 1/\nu$, then, is for $\lambda > 1/\nu$. 
In this case  $\tan \phi$ diverges as $L \rightarrow \infty$, so that  $\phi = \pi/2$.
From (\ref{ratio}), then, one finds $A_+=A_-$.
This is precisely what happens in the Ising and Potts models in $d=2$, where the coincidence of the specific-heat amplitudes 
is guaranteed by self-duality.

\vspace{1cm}

\noindent
{\emph{Note added in proof:}}
After submission of this work, we became aware of a recent extended scaling approach by Campbell and Lundow \cite{Ca11} which yields a universal amplitude ratio $A_+/A_- = 0.540(4)$. This value is also compatible with the estimate presented herein.

%

\end{document}